\documentclass{PoS}
\usepackage{amsmath,epsfig}

\newcommand\as{\alpha_{\mathrm{S}}} 
\def\ltap{\raisebox{-.6ex}{\rlap{$\,\sim\,$}} \raisebox{.4ex}{$\,<\,$}}

\title{The transverse momentum distribution of the Higgs boson at the LHC\thanks{Work done in collaboration with G.~Bozzi, S.~Catani and D.~de~Florian.}}

\ShortTitle{The transverse-momentum distribution of the Higgs boson at the LHC}

\author{Massimiliano Grazzini\\
        INFN, Sezione di Firenze, I-50019 Sesto Fiorentino, Florence, Italy\\
        E-mail: \email{grazzini@fi.infn.it}}

\abstract{We present QCD predictions for the transverse momentum ($q_T$) distribution of the Higgs boson at the LHC. At small $q_T$ the
logarithmically-enhanced terms are resummed to all orders up
to next-to-next-to-leading logarithmic accuracy.
The resummed component
is consistently matched to the next-to-leading
order calculation valid at large $q_T$. The results,
which implement the most advanced perturbative predictions
available at present for this observable,
show a good stability with respect to theoretical uncertainties.
The numerical program {\em HqT}, used to perform the calculation, is briefly discussed.
}

\FullConference{International Europhysics Conference on High Energy Physics\\
		 July 21st - 27th 2005\\
		 Lisboa, Portugal}

\begin{document}
The search for the Higgs boson
%is among the major issues
is one of the highest priorities
of the LHC physics program \cite{atlascms}.
In the last years a significant effort has been devoted to refining
the theoretical predictions for the various Higgs production
channels and the corresponding backgrounds, which are now known
to next-to-leading order accuracy (NLO) in most of the cases \cite{leshouches}.
In the case of gluon--gluon fusion, which is the main Standard Model
Higgs production channel,
even next-to-next-to leading order (NNLO) QCD corrections
have been computed, although in the large-$M_t$ 
approximation ($M_t$ being the mass of the top quark).
The result has been obtained first for the total rate \cite{NNLOtotal}, and
more recently for fully exclusive distributions \cite{NNLOdiff}.
%Predictions for more exclusive observables
%are in fact required to perform realistic studies.
Among the possible observables, an important role is played
by the transverse-momentum spectrum
of the Higgs boson, whose knowledge may help to enhance the 
statistical significance of the signal over the background.

When the transverse momentum $q_T$ of the Higgs boson is of the
order of its mass
$M_H$, the perturbative series is controlled by a small expansion
parameter, $\as(M_H^2)$, and the fixed-order prediction is reliable.
The leading order (LO) calculation \cite{Ellis:1987xu}
shows that the large-$M_t$ 
approximation works well as long
as both $M_H$ and $q_T$ are smaller than $M_t$.
In this framework, the NLO QCD corrections
have been known for some time
\cite{deFlorian:1999zd,Ravindran:2002dc,Glosser:2002gm,NNLOdiff}.

The small-$q_T$ region ($q_T\ll M_H$) is the most important, because
it is here that the bulk of events is expected. In this region
%the convergence of the fixed-order expansion is spoiled, since
the coefficients of the perturbative series in $\as(M_H^2)$ are enhanced
by powers of large logarithmic terms, $\ln^m (M_H^2/q_T^2)$. To obtain
reliable perturbative predictions, these terms have 
to be systematically resummed to all orders in $\as$ \cite{Catani:2000jh}.
%To correctly enforce transverse-momentum conservation,
%the resummation has to be carried out in $b$ space, where the impact parameter
%$b$ is the variable conjugate to $q_T$ through a Fourier transformation.
In the case of the Higgs boson,
the resummation has been explicitly worked out at
leading logarithmic (LL), next-to-leading logarithmic (NLL) 
\cite{Catani:vd}, \cite{Kauffman:cx}
and next-to-next-to-leading logarithmic (NNLL) \cite{deFlorian:2000pr} level.
The fixed-order and resummed approaches then have
to be consistently matched at intermediate values of $q_T$,
so as to avoid double counting.

In the following
we present predictions for the Higgs boson $q_T$ distribution at the LHC
within the formalism of Refs.~\cite{Catani:2000vq}--\cite{Bozzi:2005wk}.
In particular, we include the
best perturbative information that is available at present:
NNLL resummation at small $q_T$ and NLO calculation at large $q_T$.

An important feature of our formalism is that a unitarity constraint
on the total
cross section is automatically enforced, such that
the integral of the spectrum reproduces the
known fixed-order results.
More details are given in Ref.~\cite{Bozzi:2005wk}.
Other phenomenological results can be found in Ref.~\cite{recent}.

We now present quantitative results from Ref.~\cite{Bozzi:2005wk}
at NLL+LO and NNLL+NLO accuracy.
At NLL+LO (NNLL+NLO) accuracy the NLL (NNLL) resummed result is matched
to the LO (NLO) perturbative calculation valid at large $q_T$.
Our calculation is implemented in the numerical program {\tt HqT},
which can be downloaded from \cite{code}.
The code is an improved version of the
original program used in Ref.~\cite{Bozzi:2003jy},
the main difference being in the matching procedure,
which is now performed using the results of Ref.~\cite{Glosser:2002gm}.

The numerical results in Figs.~1 and 2
are obtained by choosing $M_H=125$~GeV and using 
the MRST2004 set of parton distributions \cite{Martin:2004ir}.
At NLL+LO, NLO parton densities and 
2-loop $\as$ are used, whereas at NNLL+NLO
we use NNLO parton densities 
and 3-loop $\as$.
%%====================================
\begin{figure}[htb]
\begin{center}
\begin{tabular}{c}
\epsfxsize=10truecm
\epsffile{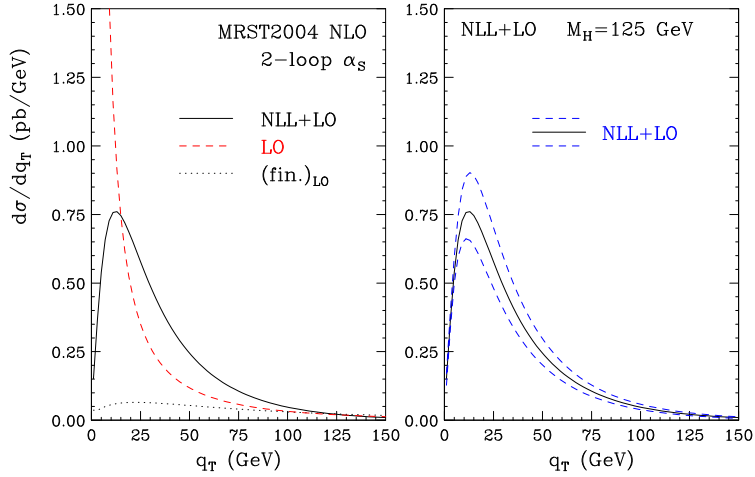}\\
\end{tabular}
\end{center}
\caption{\label{fig1}
{\em 
LHC results at NLL+LO accuracy.}}
\end{figure}
%%====================================
The NLL+LO results at the LHC are shown in Fig.~\ref{fig1}.
In the left panel, the full NLL+LO result (solid line)
is compared with the LO one (dashed line)
at the default scales $\mu_F=\mu_R=M_H$.
We see that the LO calculation diverges to $+\infty$ as $q_T\to 0$. 
The finite component, obtained through
the matching procedure, is also shown (dotted line).
The effect of the resummation, relevant below $q_T\sim 100$~GeV,
leads to a physically well defined distribution at $q_T\to 0$.
In the right panel we show the NLL+LO band obtained
by varying $\mu_F$ and $\mu_R$ simultaneously and independently
between $0.5 M_H$ and $2M_H$, imposing $0.5 \le \mu_F/\mu_R \le 2$.
The integral of the spectrum agrees with the total NLO cross section
to better than $1\%$.
%%====================================
\begin{figure}[htb]
\begin{center}
\begin{tabular}{c}
\epsfxsize=10truecm
\epsffile{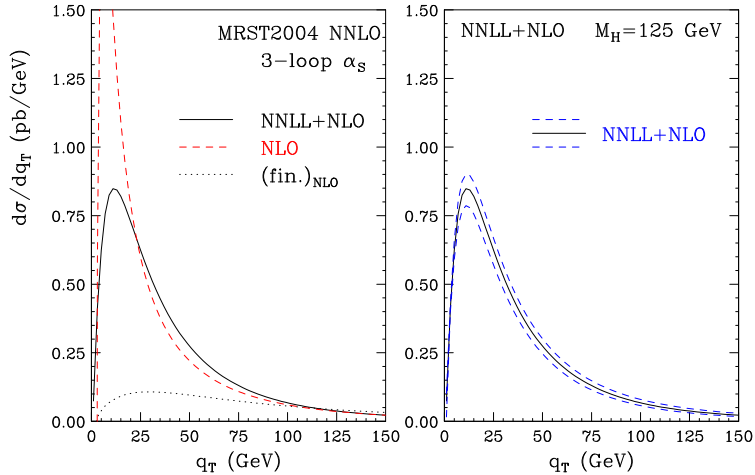}\\
\end{tabular}
\end{center}
\caption{\label{fig2}
{\em 
LHC results at NNLL+NLO accuracy. }}
\end{figure}
%%====================================
The corresponding NNLL+NLO results are shown in Fig.~\ref{fig2}.
In the left panel, the full result (solid line)
is compared with the NLO one (dashed line) at the
default scales $\mu_F=\mu_R=M_H$.
The NLO result diverges to $-\infty$ as $q_T\to 0$ and, at small values of 
$q_T$, it has an unphysical peak
that is produced by the numerical compensation of negative
leading and positive subleading logarithmic contributions.
The finite component (dotted line) vanishes smoothly as $q_T\to 0$,
showing the quality of our matching procedure.
The NNLL+NLO resummed result is slightly harder than the NLL+LO one,
and its integral is in very good agreement
with the NNLO total cross section.
The right panel of Fig.~\ref{fig2} shows the scale dependence computed as
in Fig.~\ref{fig1}.
Comparing Figs.~1 and 2, we see that the NNLL+NLO band is smaller 
than the NLL+LO one and overlaps with the latter at $q_T \ltap 100$~GeV.
This suggests a good convergence of the resummed perturbative expansion.
Other sources of perturbative uncertainty give smaller effects
\cite{Bozzi:2005wk}. 

In summary, considering the above results, the perturbative uncertainty of the NNLL+NLO spectrum is of about $10\%$ at intermediate and small $q_T$,
where the bulk of the events is concentrated.
At very small $q_T$
($q_T\ltap 10$ GeV)
non-perturbative effects should be taken into account, whereas at large $q_T$ the perturbative uncertainty increases.
Our results for the $q_T$ spectrum are thus fully consistent with
those on the total NNLO cross section \cite{NNLOtotal}.

\end{document}